\documentclass[letter,12pt]{article}%
\usepackage{amsmath,bm}
\usepackage{amsfonts}
\usepackage{amssymb}
\usepackage{graphicx}
\usepackage{rotating}
\usepackage[height=9in,left=1in,right=0.75in,bottom=1in]{geometry}
\usepackage[breaklinks=true,bookmarksopen=true,colorlinks=true,citecolor=blue]%
{hyperref}
\usepackage{color}
\usepackage{colortbl}
\usepackage{paralist}
\usepackage{amsmath}%
\providecommand{\U}[1]{\protect\rule{.1in}{.1in}}
\RequirePackage{hypernat}
\newtheorem{theorem}{Theorem}[section]

\newtheorem{assumption}{Assumption}

\newtheorem{corollary}{Corollary}[section]

\newtheorem{lemma}[theorem]{Lemma}

\catcode`~=11
\newcommand{\urltilde}{\kern -.15em\lower .7ex\hbox{~}\kern .04em}
\catcode`~=13
\def \@seccntformat#1{\csname the#1\endcsname.\quad}
\makeatother
\hypersetup{pdftitle={Escanciano}, pdfsubject={Uniform Rates for Kernel Estimators of Weakly Dependent Data}, pdfauthor={Juan Carlos Escanciano}, pdfkeywords={Uniform-in-bandwidth; Kernel estimation; Empirical process theory; Mixing}}
\begin{document}

\title{Uniform Rates for Kernel Estimators of Weakly Dependent Data}
\author{Juan Carlos Escanciano\thanks{Department of Economics, Universidad Carlos III
de Madrid, email: jescanci@eco.uc3m.es. Research funded by the Spanish
Programa de Generaci\'{o}n de Conocimiento, reference number
PGC2018-096732-B-I00. }\\Universidad Carlos III de Madrid}
\date{May 20th, 2020}
\maketitle

\begin{abstract}
This paper provides new uniform rate results for kernel estimators of
absolutely regular stationary processes that are uniform in the bandwidth and
in infinite-dimensional classes of dependent variables and regressors. Our
results are useful for establishing asymptotic theory for two-step
semiparametric estimators in time series models. We apply our results to
obtain nonparametric estimates and their rates for Expected Shortfall processes.

\vspace{2mm}

\begin{description}
\item[Keywords:] Uniform-in-bandwidth; Kernel estimation; Empirical process
theory; Mixing.

\item[\emph{JEL classification:}] C14; C22.

\item \bigskip

\item \bigskip\bigskip\bigskip\bigskip

\item \bigskip\bigskip\bigskip\bigskip

\item \bigskip

\item \bigskip

\end{description}
\end{abstract}

\pagebreak

\section{Introduction}

Kernel estimators were first introduced by Rosenblatt (1956) for density
estimation and by Nadaraya (1964) and Watson (1964) for regression estimation.
Uniform convergence for kernel estimators of weakly dependent stationary data
has been considered in a number of papers, including Bierens (1983), Liero
(1989), Roussas (1990), Peligrad (1991), Andrews (1995), Liebscher (1996),
Masry (1996), Bosq (1998), Fan and Yao (2003), Ango Nze and Doukhan (2004),
Hansen (2008), Kristenssen (2009), and Kong, Linton, and Xia (2010), among
others. In this paper we provide a general uniform rate result for kernel
estimators of absolutely regular stationary processes, where the uniformity is
in the bandwidth and over possibly infinite-dimensional classes of dependent
variables and regressors. Our results are useful for establishing asymptotic
theory for two-step semiparametric estimators in time series models.

We generalize a number of uniform-in-bandwidth results that were obtained for
independent and identically distributed observations by Einmahl and Mason
(2005) and Escanciano, Jacho-Chavez and Lewbel (2014) to the weakly dependent
stationary case. Our results complement related results given in Andrews
(1995) and Kristenssen (2009). These authors permit more heterogeneity and
different dependence concepts than ours. In contrast, we deal with unbounded
dependent variables (unlike Andrews (1995)), infinite-dimensional classes of
regressors and dependent variables, and provide uniform-in-bandwidth results
(unlike Kristenssen (2009)). We provide primitive conditions for some of the
equicontinuity assumptions required in Andrews (1995). Our conditions for
infinite-dimensional classes are relatively easy to check.

We apply empirical processes tools developed in Doukhan, Massart and Rio
(1995) to deal with the uniformity in the stochastic part of kernel
estimators, replacing the use of the celebrated Talagrand's inequality (see
Talagrand, 1994) in the work of Einmahl and Mason (2005) and Escanciano,
Jacho-Chavez and Lewbel (2014). This method of proof requires establishing
some preliminary entropy bounds for classes indexed by the bandwidth, as in
Einmahl and Mason (2005), but also over classes of dependent variables and
regressors. The entropy bounds are for a special norm introduced in Doukhan et
al. (1995), which accommodates the weak dependence structure.

We introduce notation from empirical processes theory that will be used
throughout. For a class of measurable functions $\mathcal{G}$ from
$\mathbb{R}^{p}$ to $\mathbb{R}$, let $\left\Vert \cdot\right\Vert $ be a
generic pseudo-norm on $\mathcal{G},$ defined as a norm except for the
property that $\left\Vert f\right\Vert =0$ does not necessarily imply that
$f\equiv0.$ Given two functions $l,u,$ a bracket $[l,u]$ is the set of
functions $f\in\mathcal{G}$ such that $l\leq f\leq u$. An $\varepsilon
$-bracket with respect to $\left\Vert \cdot\right\Vert $ is a bracket $[l,u]$
with $\left\Vert l-u\right\Vert \leq\varepsilon,$ $\left\Vert l\right\Vert
<\infty$ and $\left\Vert u\right\Vert <\infty$ (note that $u$ and $l$ not need
to be in $\mathcal{G}$). The \textit{covering number with bracketing}
$N_{[\cdot]}(\varepsilon,\mathcal{G},\left\Vert \cdot\right\Vert )$ is the
minimal number of $\varepsilon$-brackets with respect to $\left\Vert
\cdot\right\Vert $ needed to cover $\mathcal{G}$. These definitions are
extended to classes taking values in $\mathbb{R}^{d},$ with $d>1,$ by taking
the maximum of the bracketing numbers of the coordinate classes. Let
$\left\Vert \cdot\right\Vert _{2,\mathbb{P}}$ be the $L_{2}(\mathbb{P})$ norm,
i.e. $\left\Vert f\right\Vert _{2,\mathbb{P}}^{2}=\int f^{2}d\mathbb{P}$. When
$\mathbb{P}$ is clear from the context, we simply write $\left\Vert
\cdot\right\Vert _{2}\equiv\left\Vert \cdot\right\Vert _{2,\mathbb{P}}$. Let
$\left\vert \cdot\right\vert $ denote the Euclidean norm, i.e. $\left\vert
A\right\vert ^{2}=A^{\top}A$ ($A^{\top}$ denotes the transpose of $A$). Define
for any vector $a$ of $p$ integers the differential operator $\partial_{x}%
^{a}:=\partial^{\left\vert a\right\vert _{1}}/\partial x_{1}^{a_{1}}%
\ldots\partial x_{p}^{a_{p}},$ where $\left\vert a\right\vert _{1}:=\sum
_{t=1}^{p}a_{t}$. Let $\mathcal{S}$ be a convex set of $\mathbb{R}^{p}$, with
non-empty interior. For any smooth function $h:\mathcal{S}\subseteq
\mathbb{R}^{p}\rightarrow\mathbb{R}$ and some $\eta>0$, let \underline{$\eta$}
be the largest integer strictly smaller than $\eta$, and%
\[
\left\Vert h\right\Vert _{\infty,\eta}:=\underset{\left\vert a\right\vert
_{1}\leq\underline{\eta}}{\max}\text{ }\underset{x\in\mathcal{X}}{\sup
}\left\vert \partial_{x}^{a}h(x)\right\vert +\underset{\left\vert a\right\vert
_{1}=\underline{\eta}}{\max}\text{ }\underset{x\neq x^{\prime}}{\sup}%
\frac{\left\vert \partial_{x}^{a}h(x)-\partial_{x}^{a}h(x^{\prime})\right\vert
}{\left\vert x-x^{\prime}\right\vert ^{\eta-\underline{\eta}}}\text{.}%
\]
Further, let $C_{M}^{\eta}(\mathcal{S})$ be the set of all continuous
functions $h:\mathcal{S}\subseteq\mathbb{R}^{p}\rightarrow\mathbb{R}$ with
$\left\Vert h\right\Vert _{\infty,\eta}\leq M$. The sup norm is $\left\Vert
h\right\Vert _{\infty}:=\sup_{x\in\mathcal{S}}\left\vert h(x)\right\vert .$
Finally, throughout $C$ denotes a positive constant that may change from
expression to expression. Henceforth, we abstract from measurability issues
that may arise (see van der Vaart and Wellner (1996) for ways to deal with
lack of measurability).

\section{Uniform Rate Results}

Let $\mathcal{Z}_{n}:=\{Y_{t},X_{t}\}_{t=1}^{n}$ represent a sample of size
$n$ from a sequence of stationary and $\beta$-mixing process $Z_{t}%
=(Y_{t},X_{t}),$ where $Y_{t}$ takes values in $\mathcal{S}_{Y}\subset
\mathbb{R}^{q}$ and $X_{t}$ takes values in $\mathcal{S}_{X}\subset
\mathbb{R}^{p}$. Recall the definition of a $\beta$-mixing process. Let
$\mathcal{F}_{s}^{t}\equiv\mathcal{F}_{s}^{t}(Z_{t})$ denote the $\sigma
$-algebra generated by $\{Z_{j}$, $j=s,\ldots,t\},$ $s\leq t,$ $s,t\in
\mathbb{Z}.$ Define the $\beta$-mixing coefficients as (see, e.g., Doukhan
(1994))%
\[
\beta_{j}=\sup_{m\in\mathbb{Z}}\sup_{A\in\mathcal{F}_{j+m}^{\infty}}%
\mathbb{E}\left\vert \mathbb{P}(A|\mathcal{F}_{-\infty}^{m})-\mathbb{P}%
(A)\right\vert .
\]
Let $\Upsilon$ be a class of measurable real-valued functions of $Z_{t}$ and
let $\mathcal{W}$ be a class of measurable functions of $X_{t}$ with values in
$\mathbb{R}^{d},$ $d\leq q$. Define $\mathcal{S}_{\mathcal{W}}:=\{W(x)\in
\mathbb{R}^{d}:W\in\mathcal{W}$, $x\in\mathcal{X}_{X}\}.$ We denote by
$\psi:=(\varphi,W)$ a generic element of the set $\Psi:=\Upsilon
\times\mathcal{W}$. Let $f_{W}(w)$ denote the Lebesgue density of $W(X_{t})$
evaluated at $w$. Define the regression function $m_{\psi}(w):=\mathbb{E}%
[\varphi(Z_{t})|W\left(  X_{t}\right)  =w]$. Henceforth, we use the convention
that a function evaluated outside its support is zero. Then, an estimator for
$T_{\psi}(w):=m_{\psi}(w)f_{W}(w)$ is given by
\[
\widehat{T}_{\psi,h}(w)=\frac{1}{nh^{d}}\sum_{t=1}^{n}\varphi\left(
Z_{t}\right)  K\left(  \frac{w-W\left(  X_{t}\right)  }{h}\right)  \text{,}%
\]
where $K\left(  w\right)  =%
{\textstyle\prod\nolimits_{l=1}^{d}}
k(w_{l})$, $k\left(  \cdot\right)  $ is a kernel function, $h:=h_{n}>0$ is a
bandwidth and $w=(w_{1},\ldots,w_{d})^{\top}$. We consider the following
regularity conditions on the data generating process, kernel, bandwidth and
classes of functions.

\begin{assumption}
\label{dgp}$\{Z_{t}\}_{t\in\mathbb{Z}}$ \emph{is a strictly stationary and
absolutely regular (}$\beta$\emph{-mixing), with mixing coefficients of order
}$O(j^{-b})$\emph{, for some }$b$\emph{\ such that }$b>\delta/(\delta
-2)$\emph{, where }$2<\delta<\infty$.
\end{assumption}

\begin{assumption}
\label{compact} \emph{For }$\delta>2$\emph{ as in Assumption \ref{dgp} and
each }$1>\varepsilon>0$\emph{: (i) the class }$\Upsilon$\emph{ satisfies
}$\log N_{[\cdot]}(\varepsilon,\Upsilon,\left\Vert \cdot\right\Vert _{2})\leq
C\varepsilon^{-v_{\varphi}}$\emph{, for some }$v_{\varphi}<2,$\emph{ with an
envelope }$G(Z_{t})$\emph{\ such that }$\mathbb{E}[G(Z_{t})^{\delta}]<\infty$
\emph{and }$\sup_{w\in\mathcal{S}_{\mathcal{W}}}\mathbb{E}[G(Z_{t}%
)^{2}|W(X_{t})=w]<C$\emph{; (ii) the class }$\mathcal{W}$\emph{ is such that
(a) }$\log N(\varepsilon,\mathcal{W},\left\Vert \cdot\right\Vert _{\infty
})\leq C\varepsilon^{-v_{w}}$\emph{, for some }$v_{w}<1,$ \emph{or (b) }$\log
N(\varepsilon,\mathcal{W},\left\Vert \cdot\right\Vert _{2})\leq C\varepsilon
^{-v_{w}}$\emph{, for some }$v_{w}<1/2.$
\end{assumption}

\begin{assumption}
\label{differentiability} $T_{\psi}\in C_{M}^{r}(\mathcal{S}_{\mathcal{W}}%
)$\emph{, where }$r$\emph{ is as in Assumption \ref{kernel} below, and }%
$f_{W}(w)$\emph{ is uniformly bounded}.
\end{assumption}

\begin{assumption}
\label{kernel} \emph{The kernel function }$k\left(  t\right)  :\mathbb{R}%
\rightarrow\mathbb{R}$\emph{ is bounded, symmetric and satisfies the following
conditions: }$\int k\left(  t\right)  dt=1$\emph{, }$\int t^{l}k\left(
t\right)  dt=0$\emph{ for }$0<l<r$\emph{, and }$\int\left\vert t^{r}k\left(
t\right)  \right\vert dt<\infty$\emph{, for some }$r\geq2$\emph{. Moreover,
either }$k$ \emph{is Lipschitz and has a truncated support or }$k$\emph{ is
differentiable and satisfies }$\left\vert \partial k(t)/\partial t\right\vert
\leq C$\emph{ and for some }$v>1$\emph{, }$\left\vert \partial k(t)/\partial
t\right\vert \leq C\left\vert t\right\vert ^{-v}$\emph{ for }$\left\vert
t\right\vert >L$\emph{, }$0<L<\infty$\emph{.}
\end{assumption}

\begin{assumption}
\label{bandwidth} \emph{The possibly data-dependent bandwidth }$h$\emph{
satisfies }$\mathbb{P}(a_{n}\leq h\leq b_{n})\rightarrow1$\emph{ as
}$n\rightarrow\infty$\emph{, for deterministic sequences of positive numbers
}$a_{n}$\emph{ and }$b_{n}$\emph{ such that }$b_{n}\rightarrow0$\emph{ and
}$na_{n}^{d}\rightarrow\infty$\emph{.}
\end{assumption}

\noindent Assumption \ref{dgp} requires that observations are strictly
stationary and $\beta$-mixing, as in Doukhan, Massart and Rio (1995). As
usual, there is a tradeoff between the moments and the dependence allowed.
Assumption \ref{compact} restricts the \textquotedblleft
size\textquotedblright\ of the classes $\Upsilon$ and $\mathcal{W}$. There are
numerous examples of classes satisfying Assumption \ref{compact}, see, e.g.,
van der Vaart and Wellner (1996) and Nickl and P\"{o}tscher (2007). Note we do
not require $\mathcal{S}_{X}$ nor $\mathcal{S}_{\mathcal{W}}$ to be bounded.
Assumption \ref{differentiability} is a standard assumption used for
controlling the bias uniformly. Assumption \ref{kernel} is taken from Hansen
(2008), while Assumption \ref{bandwidth} permits data dependent bandwidths, as
in, e.g., Andrews (1995). In particular, our theory allows for plug-in
bandwidths of the form $\widehat{h}_{n}=\widehat{c}h_{n}$ with $\widehat{c}$
stochastic and $h_{n}$ a suitable deterministic sequence converging to zero as
$n\rightarrow\infty$. Andrews (1995) points out that this condition holds in
many common data dependent bandwidth selection procedures, such as
cross-validation and generalized cross-validation.

Define the rate%
\[
d_{n}:=\sqrt{\frac{1}{na_{n}^{d}}}+b_{n}^{r}.
\]

\begin{theorem}
\label{uniform_convergence}Let Assumptions \ref{dgp} -- \ref{bandwidth} hold.
Then, we have%
\begin{equation}
\underset{a_{n}\leq h\leq b_{n}}{\sup}\sup_{\psi\in\Psi}\sup_{w\in
\mathcal{S}_{\mathcal{W}}}|\widehat{T}_{\psi,h}(w)-T_{\psi}(w)|=O_{\mathbb{P}%
}(d_{n})\text{.} \label{0}%
\end{equation}

\end{theorem}

We apply the previous result to obtain rates for Nadaraya-Watson kernel
estimators. Define the kernel estimators
\begin{align*}
\widehat{m}_{\psi,h}(w)  &  :=\widehat{T}_{\psi,h}(w)/\widehat{f}%
_{W,h}(w)\text{, where}\\
\widehat{f}_{W,h}(w)  &  :=\frac{1}{nh^{d}}\sum_{t=1}^{n}K\left(
\frac{w-W\left(  X_{t}\right)  }{h}\right)  \text{.}%
\end{align*}
For a positive sequence $c_{n}$ define also%
\[
\tau_{n}=\inf_{\left\vert w\right\vert \leq c_{n},W\in\mathcal{W}}f_{W}(w)>0.
\]

\begin{corollary}
\label{uniform_convergenceF}Let Assumptions \ref{dgp}-\ref{bandwidth} and
$\tau_{n}^{-1}d_{n}=o(1)$ hold. Then, we have%
\[
\underset{l_{n}\leq h\leq u_{n}}{\sup}\underset{\psi\in\Psi}{\sup}%
\sup_{\left\vert w\right\vert \leq c_{n}}|\widehat{m}_{\psi,h}(w)-m_{\psi
}(w)|=O_{\mathbb{P}}(\tau_{n}^{-1}d_{n}).
\]

\end{corollary}

\section{Application to Conditional Expected Shortfall Processes}

There is an extensive literature on semiparametric and nonparametric
estimation of Expected Shortfall (ES). Escanciano and Mayoral (2008) review
the literature on parametric and semiparametric estimation of ES and provide a
unified approach; see also Nadarajah, Zhang and Chan (2014). Nonparametric
estimation of Conditional ES (CES) has been studied by Scaillet (2004). He
proposed a kernel estimator for the quantity
\[
CES_{a,p}:=\mathbb{E}[-a^{\top}Y_{t}|-a^{\top}Y_{t}>VaR(a,p)],
\]
where the vector $a$ are portfolio weights, $a\in\mathcal{A}\subseteq
\{a\in\mathbb{R}^{q}:\left\vert a\right\vert =1\},$ and $VaR(a,p)$ is the
$p-th$ Value-at-Risk (VaR), $p\in(0,1),$ defined as%
\[
\mathbb{P}\left(  -a^{\top}Y_{t}>VaR(a,p)\right)  =p.
\]
We introduce covariates and study nonparametric estimation of
\[
CES_{a,b,c}(w):=\mathbb{E}[-a^{\top}Y_{t}|-a^{\top}Y_{t}>c(X),b^{\top}%
X_{t}=w],
\]
as a process in $(a,b,c,w).$ Portfolio weights are often estimated. The
motivation to consider $b^{\top}X_{t}$ is to reduce the dimensionality of the
conditioning set. The motivation to consider a function $c(X)$ is to be able
to obtain rates when a plugging estimator for the conditional VaR is
considered. Fully nonparametric estimators for ES with covariates are proposed
in Scaillet (2005), Cai and Wang (2008), and Linton and Xiao (2013). An
application of the smoothed ES estimator of Scaillet (2004) with generated
variables is given in Brownlees and Engle (2016).

To study $CES_{a,b,c},$ we use that
\[
CES_{a,b,c}(w)=\frac{\mathbb{E}[\varphi_{1}\left(  Z_{t}\right)  |W\left(
X_{t}\right)  =w]}{\mathbb{E}[\varphi_{2}\left(  Z_{t}\right)  |W\left(
X_{t}\right)  =w]},
\]
where $\varphi_{1}\in\mathcal{F}_{1},$ $\varphi_{2}\in\mathcal{F}_{2}$ and
$W\in\mathcal{W},$ with%
\begin{align*}
\mathcal{F}_{1}  &  =\left\{  (y,x)\rightarrow-a^{\top}y1(-a^{\top
}y>c(x)):a\in\mathcal{A},c\in\mathcal{C}\right\} \\
\mathcal{F}_{2}  &  =\left\{  (y,x)\rightarrow1(-a^{\top}y>c(x)):a\in
\mathcal{A},c\in\mathcal{C}\right\} \\
\mathcal{W}  &  =\left\{  x\rightarrow b^{\top}x:b\in\mathcal{B}%
\subset\mathbb{R}^{p}\right\}  .
\end{align*}
Here $1(E)$ is the indicator function of the event $E,$ which equals one if
$E$ is true and zero otherwise. A kernel estimator for $CES_{a,b,c}$ is then
\[
\widehat{CES}_{a,b,c}(w)=\frac{\frac{1}{nh}\sum_{t=1}^{n}\varphi_{1}\left(
Z_{t}\right)  K\left(  \frac{w-W\left(  X_{t}\right)  }{h}\right)  }{\frac
{1}{nh}\sum_{t=1}^{n}\varphi_{2}\left(  Z_{t}\right)  K\left(  \frac
{w-W\left(  X_{t}\right)  }{h}\right)  },
\]
To apply our previous results, write $CES_{a,b,c}(w)$ and its estimator as
indexed by $\psi:=(\varphi_{1},\varphi_{2},W)\in\Psi:=\mathcal{F}_{1}%
\times\mathcal{F}_{2}\times\mathcal{W}.$ Thus, we write $CES_{a,b,c}%
(w)=CES_{\psi}(w).$ Define the functions $m_{\psi_{j}}(w):=\mathbb{E}%
[\varphi_{j}(Z_{t})|W\left(  X_{t}\right)  =w]$ and $T_{\psi_{j}}%
(w):=m_{\psi_{j}}(w)f_{W}(w)$ for $j=1,2$. Let $\lambda_{\min}(A)$ and
$\lambda_{\max}(A)$ denote the minimum and maximum eigenvalue for a positive
definite symmetric matrix $A.$ Then, consider the following assumptions:

\begin{assumption}
\label{classes} \emph{(i)} $\mathbb{E}[\left\vert Y_{t}\right\vert
^{2}]<\infty$ \emph{and uniformly in }$b\in\mathbb{R}^{p}:0<\lambda_{\min
}(\mathbb{E}[Y_{t}Y_{t}^{\top}|b^{\top}X_{t}])\leq\lambda_{\max}%
(\mathbb{E}[Y_{t}Y_{t}^{\top}|b^{\top}X_{t}])<C$\emph{ a.s.; (ii) the class
}$\mathcal{C}$\emph{ is such that }$\log N(\varepsilon,\mathcal{C},\left\Vert
\cdot\right\Vert _{\infty})\leq C\varepsilon^{-v_{c}}$\emph{, for some }%
$v_{c}<1;$ \emph{(iii)} $\mathcal{B}$ \emph{is compact and} $\mathbb{E}%
[\left\vert X_{t}\right\vert ^{2}]<\infty.$
\end{assumption}

\begin{assumption}
\label{differentiability2} $T_{\psi_{j}}\in C_{M}^{r}(\mathcal{S}%
_{\mathcal{W}})$\emph{, where }$r$\emph{ is as in Assumption \ref{kernel}, and
the conditional and marginal densities of }$a^{\top}Y_{t}$ \emph{given
}$b^{\top}X_{t}$\emph{ and }$b^{\top}X_{t},$\emph{ respectively, are}
\emph{uniformly bounded (in }$a\in\mathcal{A}$\emph{ and }$b\in\mathcal{B}%
)$\emph{.}
\end{assumption}

Define the rate%
\[
d_{n}:=\sqrt{\frac{1}{na_{n}}}+b_{n}^{r}.
\]

For a positive sequence $c_{n}$ define also%
\[
\tau_{n}=\inf_{\left\vert w\right\vert \leq c_{n},\psi_{2}\in\mathcal{F}_{2}%
}T_{\psi_{2}}(w)>0.
\]

\begin{theorem}
\label{ES}Let Assumptions \ref{dgp}, \ref{kernel}, \ref{bandwidth},
\ref{classes}, \ref{differentiability2} and $\tau_{n}^{-1}d_{n}=o(1)$ hold.
Then, we have%
\[
\underset{l_{n}\leq h\leq u_{n}}{\sup}\underset{\psi\in\Psi}{\sup}%
\sup_{\left\vert w\right\vert \leq c_{n}}|\widehat{CES}_{\psi}(w)-CES_{\psi
}(w)|=O_{\mathbb{P}}(\tau_{n}^{-1}d_{n}).
\]

\end{theorem}

\section{Proofs}

\noindent\textbf{Proof of Theorem \ref{uniform_convergence}}: Write
\begin{align*}
\sup|\widehat{T}_{\psi,h}(w)-T_{\psi}(w)|  &  \leq\sup\left\vert
\widehat{T}_{\psi,h}(w)-\mathbb{E}\left[  \widehat{T}_{\psi,h}(w)\right]
\right\vert +\sup\left\vert \mathbb{E}\left[  \widehat{T}_{\psi,h}(w)\right]
-T_{\psi}(w)\right\vert \\
&  \equiv S_{n}+B_{n}\text{,}%
\end{align*}
where henceforth the $\sup$ is over the set in the left hand side of
(\ref{0}). We start investigating the stochastic part $S_{n}$. Define the
product class of functions $\mathcal{G}_{0}:=\mathcal{K}_{0}\cdot \Upsilon$,
where%
\[
\mathcal{K}_{0}=\left\{  x\rightarrow K\left(  \frac{w-W\left(  x\right)  }%
{h}\right)  :w\in\mathcal{S}_{\mathcal{W}},W\in\mathcal{W},h\in(0,1]\right\}
.
\]
From the boundedness of the kernel, and the squared integrable envelope in
Assumption \ref{compact} it is straightforward to prove that, for some
positive constant $C$,%
\begin{equation}
N_{[\cdot]}(\varepsilon,\mathcal{G}_{0},\left\Vert \cdot\right\Vert _{2})\leq
N_{[\cdot]}\left(  C\varepsilon,\mathcal{K}_{0},\left\Vert \cdot\right\Vert
_{2}\right)  \times N_{[\cdot]}\left(  C\varepsilon,\Upsilon,\left\Vert
\cdot\right\Vert _{2}\right)  \text{.} \label{brack1}%
\end{equation}
By Lemma B.3 in Escanciano, Jacho-Ch\'{a}vez and Lewbel (2014) $\mathcal{K}%
_{0}$ satisfies
\[
N_{[\cdot]}\left(  C\varepsilon,\mathcal{K}_{0},\left\Vert \cdot\right\Vert
_{2}\right)  \leq C\varepsilon^{-\alpha_{K}}N(\varepsilon^{2},\mathcal{W}%
,\left\Vert \cdot\right\Vert _{\infty}),
\]
for some $\alpha_{K}\geq1.$ An by Lemma A1 in Escanciano and Zhu (2015)
\[
N_{[\cdot]}\left(  C\varepsilon,\mathcal{K}_{0},\left\Vert \cdot\right\Vert
_{2}\right)  \leq C\varepsilon^{-\alpha_{K}}N(\varepsilon^{4},\mathcal{W}%
,\left\Vert \cdot\right\Vert _{2}).
\]
An inspection of the proof of these two Lemmas reveals that $\mathcal{S}%
_{\mathcal{W}}$ could be unbounded. Hence, by our assumptions on the classes
$\Upsilon$ and $\mathcal{W}$, we obtain that $\log N_{[\cdot]}(\varepsilon
,\mathcal{G}_{0},\left\Vert \cdot\right\Vert _{2})\leq C\varepsilon^{-v}$, for
some $v<2$. Define the norm
\[
\Vert f\Vert_{2,\beta}^{2}=\int_{0}^{1}\beta^{-1}(u)Q_{f}^{2}(u)du,
\]
where $\beta^{-1}$ is the inverse cadlag of the decreasing function
$u\rightarrow\beta_{\lfloor u\rfloor}$ ($\lfloor u\rfloor$ being the integer
part of $u$, and $\beta_{t}$ being the mixing coefficient) and $Q_{f}$ is the
inverse cadlag of the tail function $u\rightarrow\mathbb{P}(|f|>u)$ (see
Doukhan, Massart and Rio 1995). Note that
\[
\mathbb{P}\left(  \left\vert f-g\right\vert >z\right)  \leq\frac
{\mathbb{E}[\left\vert f-g\right\vert ^{2}]}{z^{2}}%
\]
and hence, for an $\sqrt{b}\varepsilon-$bracket $[f,g]$ wrt $\Vert\cdot
\Vert_{2}$
\[
\Vert f-g\Vert_{2,\beta}^{2}\leq\int_{0}^{1}\beta^{-1}(u)\frac{b\varepsilon
^{2}}{u}du\leq b\varepsilon^{2}\int_{0}^{1}u^{b-1}du=\varepsilon^{2}.
\]
Therefore,
\begin{align*}
\log N_{[\cdot]}\left(  \varepsilon,\mathcal{G}_{0},\Vert\cdot\Vert_{2,\beta
}\right)   &  \leq\log N_{[\cdot]}\left(  \sqrt{b}\varepsilon,\mathcal{G}%
_{0},\left\Vert \cdot\right\Vert _{2}\right) \\
&  \leq C\varepsilon^{-v}.
\end{align*}
Theorem 3 in Doukhan, Massart and Rio (1995) applied to the class
$\mathcal{G}_{0}$ then implies
\[
\underset{l_{n}\leq h\leq u_{n}}{\sup}\underset{\psi\in\Psi}{\sup}\left\vert
\widehat{T}_{h}(\psi)-\mathbb{E}\left[  \widehat{T}_{h}(\psi)\right]
\right\vert =O_{\mathbb{P}}\left(  \sqrt{\frac{1}{na_{n}^{d}}}\right)  ,
\]
provided $\Vert f\Vert_{2,\beta}\leq Ch^{d/2}$ for all $f\in\mathcal{G}_{0}$.
But by Assumption \ref{kernel} and Pollard (1984, pg. 36)
\begin{align*}
\mathbb{P}\left(  \left\vert f\right\vert >z\right)   &  \leq\frac
{\mathbb{E}[\left\vert f\right\vert ^{2}]}{z^{2}}\\
&  \leq\frac{Ch^{d}}{z^{2}},
\end{align*}
where have used $\sup_{w\in\mathcal{S}_{\mathcal{W}}}\mathbb{E}[G(Z_{t}%
)^{2}|W(X_{t})=w]<C$ and the bounded density and kernel assumption. Hence,
\[
\Vert f\Vert_{2,\beta}^{2}\leq\int_{0}^{1}\beta^{-1}(u)\frac{Ch^{d}}{u}du\leq
Ch^{d}\int_{0}^{1}u^{b-1}du=\frac{Ch^{d}}{b},
\]
where the latter inequality follows from Assumption \ref{dgp}.

We now study the bias part $B_{n}$. By a multivariate Taylor expansion
\[
T_{\psi}(w+uh)=%
{\textstyle\sum\limits_{\left\vert \alpha\right\vert _{1}<r-1}}
\frac{\partial_{w}^{\alpha}T_{\psi}(w)}{\alpha!}(uh)^{\alpha}+%
{\textstyle\sum\limits_{\left\vert \alpha\right\vert _{1}=r-1}}
\frac{R_{\alpha}(w+uh)}{\alpha!}(uh)^{\alpha},
\]
where the remainder satisfies
\[
R_{\alpha}(w+uh)=(r-1)%
{\textstyle\int\limits_{0}^{1}}
(1-\tau)^{r-2}\partial_{w}^{\alpha}T_{\psi}(w+\tau uh)d\tau.
\]
Since $T_{\psi}\in C_{M}^{r}(\mathcal{S}_{\mathcal{W}}),$
\begin{align*}
\left\vert R_{\alpha}(w+uh)-\partial_{w}^{\alpha}T_{\psi}(w)\right\vert  &
\leq(r-1)\left\vert
{\textstyle\int\limits_{0}^{1}}
(1-\tau)^{r-2}\left[  \partial_{w}^{\alpha}T_{\psi}(w+\tau uh)-\partial
_{w}^{\alpha}T_{\psi}(w)\right]  d\tau\right\vert \\
&  \leq(r-1)M\left\vert
{\textstyle\int\limits_{0}^{1}}
(1-\tau)^{r-2}\left\vert \tau uh\right\vert d\tau\right\vert \\
&  \leq M\left\vert uh\right\vert .
\end{align*}
Thus, by a standard change of variables and Assumption \ref{kernel}%
\begin{align*}
\left\vert \mathbb{E}\left[  \widehat{T}_{\psi,h}(w)\right]  -T_{\psi
}(w)\right\vert  &  =\left\vert \int\left[  T_{\psi}(w+uh)-T_{\psi}(w)\right]
K(u)du\right\vert \\
&  =\left\vert \int%
{\textstyle\sum\limits_{\left\vert \alpha\right\vert _{1}=r-1}}
\frac{1}{\alpha!}\left[  R_{\alpha}(w+uh)-\partial_{w}^{\alpha}T_{\psi
}(w)\right]  \left(  uh\right)  ^{\alpha}K(u)du\right\vert \\
&  \leq h^{r}%
{\textstyle\sum\limits_{\left\vert \alpha\right\vert _{1}=r-1}}
\frac{M}{\alpha!}\int\left\vert u\cdot K(u)\right\vert ^{r}du.
\end{align*}
Hence,%
\[
\sup\left\vert \mathbb{E}\left[  \widehat{T}_{\psi,h}(w)\right]  -T_{\psi
}(w)\right\vert =O\left(  b_{n}^{r}\right)  .
\]
\hfill\emph{Q.E.D.}\bigskip

\noindent\textbf{Proof of Corollary \ref{uniform_convergenceF}}: From Theorem
\ref{uniform_convergence}%
\[
\underset{l_{n}\leq h\leq u_{n}}{\sup}\underset{\psi\in\Psi}{\sup}%
\sup_{\left\vert w\right\vert \leq c_{n}}|\widehat{T}_{\psi,h}(w)-T_{\psi
}(w)|=O_{\mathbb{P}}(d_{n})
\]
and%
\[
\underset{l_{n}\leq h\leq u_{n}}{\sup}\underset{W\in\mathcal{W}}{\sup}%
\sup_{\left\vert w\right\vert \leq c_{n}}|\widehat{f}_{W,h}(w)-f_{W}%
(w)|=O_{\mathbb{P}}(d_{n}).
\]
Therefore
\[
\underset{l_{n}\leq h\leq u_{n}}{\sup}\underset{W\in\mathcal{W}}{\sup}%
\sup_{\left\vert w\right\vert \leq c_{n}}\left\vert \frac{\widehat{f}%
_{W,h}(w)-f_{W}(w)}{f_{W}(w)}\right\vert =O_{\mathbb{P}}(\tau_{n}^{-1}d_{n})
\]
and%
\[
\underset{l_{n}\leq h\leq u_{n}}{\sup}\underset{W\in\mathcal{W}}{\sup}%
\sup_{\left\vert w\right\vert \leq c_{n}}\left\vert \frac{\widehat{T}_{\psi
,h}(w)-T_{\psi}(w)}{f_{W}(w)}\right\vert =O_{\mathbb{P}}(\tau_{n}^{-1}d_{n}).
\]
Thus, uniformly in $l_{n}\leq h\leq u_{n},$ $\psi\in\Psi$ and $\left\vert
w\right\vert \leq c_{n}$
\[
\widehat{m}_{\psi,h}(w)=\frac{\widehat{T}_{\psi,h}(w)/f_{W}(w)}{\widehat{f}%
_{W,h}(w)/f_{W}(w)}=\frac{m_{\psi}(w)+O_{\mathbb{P}}(\tau_{n}^{-1}d_{n}%
)}{1+O_{\mathbb{P}}(\tau_{n}^{-1}d_{n})}=m_{\psi}(w)+O_{\mathbb{P}}(\tau
_{n}^{-1}d_{n}).
\]
\emph{Q.E.D.}

The following result is well-known in empirical processes theory. Define the
generic class of measurable functions $\mathcal{F}:=\{x\rightarrow
m(x,\theta,h):\theta\in\Theta,h\in\mathcal{H}\}$, where $\Theta$ and
$\mathcal{H}$ are endowed with the pseudo-norms $\left\vert \cdot\right\vert
_{\Theta}$ and $\left\vert \cdot\right\vert _{\mathcal{H}}$, respectively.

\begin{lemma}
\label{L2cont}(Pollard; Chen, Linton and Van Keilegom) Assume that for all
$(\theta_{0},h_{0})\in\Theta\times\mathcal{H}$, $m(z,\theta,h)$ is locally
uniformly $||\cdot||_{2}$ continuous, in the sense that
\[
\mathbb{E}\left[  \sup_{\theta:\left\vert \theta_{0}-\theta\right\vert
_{\Theta}<\delta,h:\left\vert h_{0}-h\right\vert _{\mathcal{H}}<\delta
}\left\vert m(Z,\theta,h)-m(Z,\theta_{0},h_{0})\right\vert ^{2}\right]  \leq
C\delta^{s}\text{,}%
\]
for all sufficiently small $\delta>0$, some constant $s\in(0,2]$ and $C>0$.
Then,
\[
N_{[\cdot]}(\varepsilon,\mathcal{F},\left\Vert \cdot\right\Vert _{2})\leq
N\left(  \left(  \frac{\varepsilon}{2C}\right)  ^{2/s},\Theta,\left\vert
\cdot\right\vert _{\Theta}\right)  \times N\left(  \left(  \frac{\varepsilon
}{2C}\right)  ^{2/s},\mathcal{H},\left\vert \cdot\right\vert _{\mathcal{H}%
}\right)  \text{.}%
\]

\end{lemma}

\noindent\textbf{Proof of Theorem \ref{ES}}: The proof proceeds as in
Corollary \ref{uniform_convergenceF} after checking the conditions of Theorem
\ref{uniform_convergence} to obtain, for $j=1,2,$%
\[
\underset{l_{n}\leq h\leq u_{n}}{\sup}\underset{\psi_{j}\in\Psi}{\sup}%
\sup_{\left\vert w\right\vert \leq c_{n}}|\widehat{T}_{\psi_{j},h}%
(w)-T_{\psi_{j}}(w)|=O_{\mathbb{P}}(d_{n}),
\]
where%
\[
\widehat{T}_{\psi_{j},h}(w)=\frac{1}{nh}\sum_{t=1}^{n}\varphi_{j}\left(
Z_{t}\right)  K\left(  \frac{w-W\left(  X_{t}\right)  }{h}\right)  .
\]
To verify Assumption \ref{differentiability} with $\Upsilon=\mathcal{F}_{1}$
we apply Lemma \ref{L2cont} with $z=(y,x),$
\[
m(z,\theta,h)=-\theta^{\top}y1(-\theta^{\top}y>h(x))
\]
$\Theta=\mathcal{A}$ and $\mathcal{H}=\mathcal{C}$ with $\left\vert
\cdot\right\vert _{\mathcal{H}}=\left\Vert \cdot\right\Vert _{\infty}.$ We
then obtain by triangle inequality%
\begin{align*}
&  \mathbb{E}\left[  \sup_{\theta:\left\vert \theta_{0}-\theta\right\vert
_{\Theta}<\delta,h:\left\vert h_{0}-h\right\vert _{\mathcal{H}}<\delta
}\left\vert m(Z,\theta,h)-m(Z,\theta_{0},h_{0})\right\vert ^{2}\right]  \\
&  \leq2\mathbb{E}\left[  \sup_{\theta:\left\vert \theta_{0}-\theta\right\vert
_{\Theta}<\delta,h:\left\vert h_{0}-h\right\vert _{\mathcal{H}}<\delta
}\left\vert m(Z,\theta,h)-m(Z,\theta_{0},h)\right\vert ^{2}\right]  \\
&  +2\mathbb{E}\left[  \sup_{\theta:\left\vert \theta_{0}-\theta\right\vert
_{\Theta}<\delta,h:\left\vert h_{0}-h\right\vert _{\mathcal{H}}<\delta
}\left\vert m(Z,\theta_{0},h)-m(Z,\theta_{0},h_{0})\right\vert ^{2}\right]  \\
&  \leq2\delta^{2}\mathbb{E}\left[  \left\vert Y_{t}\right\vert ^{2}\right]
+C\delta.
\end{align*}
where the last inequality uses that $\left\vert m(z,\theta,h)-m(z,\theta
_{0},h)\right\vert \leq\left\vert \theta-\theta_{0}\right\vert \left\vert
y\right\vert $ and
\begin{align*}
&  \mathbb{E}\left[  \sup_{\theta:\left\vert \theta_{0}-\theta\right\vert
_{\Theta}<\delta,h:\left\vert h_{0}-h\right\vert _{\mathcal{H}}<\delta
}\left\vert m(Z,\theta_{0},h)-m(Z,\theta_{0},h_{0})\right\vert ^{2}\right]  \\
&  \leq\mathbb{E}\left[  \left(  \theta_{0}^{\top}Y_{t}\right)  ^{2}%
1(h_{0}(X_{t})-\delta<-\theta_{0}^{\top}Y_{t}<h_{0}(X_{t})+\delta)\right]  \\
&  \leq C\delta
\end{align*}
by Assumption \ref{differentiability2}. Then, Lemma \ref{L2cont} implies%
\begin{align*}
N_{[\cdot]}(\varepsilon,\Upsilon,\left\Vert \cdot\right\Vert _{2}) &  \leq
N\left(  \left(  \frac{\varepsilon}{2C}\right)  ^{2},\Theta,\left\vert
\cdot\right\vert _{\Theta}\right)  \times N\left(  \left(  \frac{\varepsilon
}{2C}\right)  ^{2},\mathcal{C},\left\Vert \cdot\right\Vert _{\infty}\right)
\\
&  \leq C\varepsilon^{-v_{\varphi}},
\end{align*}
with $v_{\varphi}=2v_{c}<2.$ The entropy condition on $\mathcal{W}$ in
Assumption \ref{differentiability}(ii-b) follows from the compactness of
$\mathcal{B}$ and $\mathbb{E}\left[  \left\vert X_{t}\right\vert ^{2}\right]
<\infty.$ This concludes the verification of Assumption
\ref{differentiability}. The same arguments apply to $\Upsilon=\mathcal{F}%
_{2}.$ Conclude as in Corollary \ref{uniform_convergenceF}. \emph{Q.E.D.}

\end{document}